\documentclass[preprint]{imsartp}

\usepackage[left=1in,right=1in,top=1in,bottom=1in]{geometry}

\RequirePackage{amsthm,amsmath,amsfonts,amssymb}
\RequirePackage[authoryear]{natbib}
\RequirePackage[colorlinks,citecolor=blue,urlcolor=blue]{hyperref}
\RequirePackage{graphicx}

\startlocaldefs
\theoremstyle{plain}

\newtheorem{theorem}{Theorem}[section]

\theoremstyle{remark}

\usepackage{amsmath,amssymb,graphicx,color,amsthm}
\usepackage{tikz}
\usepackage{dsfont}
\usepackage{mathtools}
\usetikzlibrary{shapes.geometric, arrows}
\usepackage[english]{babel}
\usepackage[utf8]{inputenc}
\usepackage{mathrsfs}  
\usepackage{afterpage}
\usepackage{mathtools}
\usepackage{amsfonts}
\usepackage[colorinlistoftodos]{todonotes}
\usepackage{enumerate}
\usepackage{hyperref}
\usetikzlibrary{automata,positioning}
\usetikzlibrary{decorations.pathreplacing}
\usepackage{algorithm, algpseudocode}
\newcommand{\excise}[1]{}

\newcommand\numberthis{\addtocounter{equation}{1}\tag{\theequation}}

\newcommand\RR{\mathbb{R}}

\newcommand\EE{\mathbb{E}}

\newcommand{\tr}{\operatorname{tr}}

\newtheorem*{example*}{Example}

\newtheorem{corollary}{Corollary}

\DeclareMathOperator\var{var}

\DeclareMathOperator\argmax{argmax}

\DeclareMathOperator\Id{I}

\DeclareMathOperator\GP{GP}

\DeclarePairedDelimiterX{\infdivx}[2]{(}{)}{%
	#1\;\delimsize\|\;#2%
}

\newcommand{\RNum}[1]{\uppercase\expandafter{\romannumeral #1\relax}}

\xdefinecolor{dukeblue}{rgb}{0.004,0.129,0.412}

\endlocaldefs

\begin{document}

\begin{frontmatter}
\title{Contrastive linear regression}
\runtitle{Contrastive linear regression}

\begin{aug}
\author[A]{\fnms{Boyang}~\snm{Zhang}\ead[label=e1]{zhangby@stanford.edu}},
\author[B]{\fnms{Sarah}~\snm{Nyquist}\ead[label=e2]{sarah.nyquist@gladstone.ucsf.edu}}
\author[C]{\fnms{Andrew}~\snm{Jones}\ead[label=e3]{aj13@princeton.edu}}
\author[B,D]{\fnms{Barbara E.}~\snm{Engelhardt}\ead[label=e4]{barbara.engelhardt@gladstone.ucsf.edu}}
\and
\author[E]{\fnms{Didong}~\snm{Li}\ead[label=e5]{didongli@unc.edu}}
\address[A]{Department of Genetics,
Stanford University\printead[presep={,\ }]{e1}}

\address[B]{Gladstone Institutes\printead[presep={,\ }]{e2,e4}}

\address[C]{Department of Computer Science,
Princeton University\printead[presep={,\ }]{e3}}

\address[D]{Department of Biomedical Data Science, Stanford University \printead[presep={,\ }]{e4}}

\address[E]{Department of Biostatistics,
University or North Carolina at Chapel Hill\printead[presep={,\ }]{e5}}

\end{aug}

\begin{abstract}
Contrastive dimension reduction methods have been developed for case-control study data to identify variation that is enriched in the foreground (case) data $X$ relative to the background (control) data $Y$. Here, we develop contrastive regression for the setting when there is a response variable $r$ associated with each foreground observation. This situation occurs frequently when, for example, the unaffected controls do not have a disease grade or intervention dosage but the affected cases have a disease grade or intervention dosage, as in autism severity, solid tumors stages, polyp sizes, or warfarin dosages. Our contrastive regression model captures shared low-dimensional variation between the predictors in the cases and control groups, and then explains the case-specific response variables through the variance that remains in the predictors after shared variation is removed. We show that, in one single-nucleus RNA sequencing dataset on autism severity in postmortem brain samples from donors with and without autism and in another single-cell RNA sequencing dataset on cellular differentiation in chronic rhinosinusitis with and without nasal polyps, our contrastive linear regression performs feature ranking and identifies biologically-informative predictors associated with response that cannot be identified using other approaches.
\end{abstract}


\end{frontmatter}

\section{Introduction}

Case-control studies are a common and powerful research design in biomedical investigations, primarily due to their ability to establish associations between risk factors or exposures and disease outcomes. These studies are particularly advantageous for investigating rare diseases, as they begin with a group of individuals with the treatment or disease (cases) and compare them to a group without the treatment or disease (controls). By selecting cases and controls from the same population, researchers can efficiently assess the relationship between each high-dimensional observation and the treatment or disease of interest. Case-control studies are also relatively quicker and more cost-effective than other study designs, such as cohort studies or prospective studies, making them an attractive choice for researchers.

The importance of analyzing data from case-control studies lies in the valuable insights they provide in understanding disease etiology, disease progression, treatment response, survival rates, and the occurrence of adverse events. Accurate prediction of these outcome variables is essential for tailoring personalized treatment plans; understanding which features are most predictive of outcome further advances our knowledge of disease mechanisms and suggests targeted interventions. A standard workflow for the analysis of case-control study data includes performing dimension reduction and prediction methods to analyze case-control study data. The results allow researchers to discover biologically-meaningful patterns and associations, ultimately leading to a better understanding of diseases and contributing to advances in biomedical research.

Contrastive dimension reduction methods characterize low-dimensional variation that is enriched in a foreground dataset $X$ relative to a background dataset $Y$. Here, we consider an extension of this setting where there is also a response variable $r$ associated with each foreground observation. In this context, we consider fitting a regression function to predict the response variable from the foreground data $X$ while controlling for shared variation in the foreground $X$ and background data $Y$.

\subsection{Dimension reduction} 
Dimension reduction is a crucial technique for the analysis of high-dimensional biomedical data, as it enables researchers to uncover meaningful patterns, structures, and relationships within complex datasets. The main objective of dimension reduction is to reduce the dimension of the data while preserving essential information, which facilitates data visualization, improves statistical inference, and enhances computational efficiency. Over the years, various dimension reduction methods have been developed and applied to various types of biomedical data, including genomic, proteomic, and imaging data. 

Linear dimension reduction techniques, such as principal component analysis (PCA,~\cite{abdi2010principal}) and multidimensional scaling (MDS,~\cite{carroll1998multidimensional}), have been widely used in biomedical research. PCA identifies the directions of maximum variance in the data and has been applied to numerous biomedical applications, such as gene expression analysis and neuroimaging. MDS aims to preserve the pairwise distances between data points in a lower-dimensional space and has been used for the visualization of genetic distances and protein structure analysis. Nonlinear dimension reduction, such as t-Distributed Stochastic Neighbor Embedding (t-SNE,~\cite{van2008visualizing}) and Uniform Manifold Approximation and Projection (UMAP,~\cite{becht2019dimensionality}), have been developed to capture complex, nonlinear relationships in high-dimensional data. These methods have been popular in scRNA-seq data analysis to visualize cell populations and identify cell types. Many more diverse methods are available.

Despite the variety of dimension reduction methods, their common objective is to find low-dimensional representations of high-dimensional data while preserving meaningful information. In case-control studies, the information of interest is often exclusive to the case group. Consequently, traditional dimension reduction methods may not adequately capture the low-dimensional information that is unique to the case group, highlighting the need for tailored approaches for these studies.

Contrastive dimension reduction (CDR) methods are specifically designed for case-control studies, aiming to identify information specific to the treatment group. Contrastive latent variable models (CLVMs,~\cite{severson2019unsupervised}), contrastive PCA (CPCA,~\cite{abid2018exploring}), probabilistic contrastive PCA (PCPCA,~\cite{li2020probabilistic}), and the contrastive Poisson latent variable model (CPLVM~\cite{jones2022contrastive}) are popular existing CDR methods to find a low-dimensional representation of the treatment group data while accounting for the structure of the control group data.
In particular, PCPCA is a probabilistic extension of CPCA that uses a generative modeling framework to capture variation unique to the treatment group data~\citep{li2020probabilistic}. This method enables a structured modeling of the data, allowing for the incorporation of prior knowledge and the estimation of uncertainty in the low-dimensional representation. CPLVM is a count-based CDR technique specifically designed for high-dimensional count data, such as gene expression data~\cite{jones2022contrastive}. This method models the count data using a Poisson data likelihood and seeks to identify covariation unique to the treatment group data while accounting for covariation patterns in the control group data. 

CDR methods have shown promise in various case-control study applications, enabling researchers to identify and characterize the patterns and structures unique to the treatment group. By focusing on the differences between the treatment and control groups, these methods can provide more targeted insights into the underlying biology of the disease or condition under investigation, ultimately contributing to the development of more effective treatments and interventions because all of the normal variation and standard processes found in the controls have been effectively subtracted from the results. 

Despite the promise of CDR methods, they all operate within the unsupervised regime, which means that they do not include outcome variables. In biomedical case-control studies, incorporating outcome variables is often of critical importance, as they may capture the severity or progression of the disease or specific rare events. For instance, identifying variability in gene expression associated uniquely with variability in patient survival rates, disease progression, or treatment response cannot be done without access to control data or in a CDR setting. As a result, there is an urgent need for contrastive regression models to analyze these complex and ubiquitous data. 

\subsection{Regression: Prediction, association testing, and variable selection} 
Regression analysis is a powerful statistical tool for studying relationships between variables. It is widely used across various domains to make predictions, test for associations, perform variable selection, and inform decision-making processes. In particular, linear regression models the relationship between outcome variables and covariates by fitting a linear equation to the observed data. It is useful for situations where the relationship between variables can be approximated by a straight line, but it may not be suitable when the relationship is nonlinear. Gaussian process regression (GPR) is a flexible, nonparametric method that models smooth nonlinear relationships between variables. GPR assumes that the dependent variable is a realization of a Gaussian process, which is defined by its mean function and covariance function. This approach allows for the incorporation of prior knowledge about the relationship between variables and provides a probabilistic framework for making predictions, thus allowing for the quantification of uncertainty in the predicted values. 

While linear and Gaussian process regression have been successful in a wide range of applications, their effectiveness in predicting outcome variables for case-control studies may be limited due to their inability to explicitly account for the unique low-dimensional structure enriched in the treatment group. Moreover, it is unclear how to address the (often) missing response variables in the control group. Therefore, there is a need for regression models that can make accurate and informative predictions in the context of case-control study data with response variables only available for the cases.

In this work, we develop a contrastive linear regression model (CLRM) tailored for case-control studies. Our approach consists of several key steps. (1) We construct latent variable models to capture the case-control structure, with one loading matrix representing the shared information between both groups and the other loading matrix representing the information unique to the treatment group. (2) The outcome variable is modeled as a linear function of the low-dimensional representation unique to the treatment group, which allows us to directly link the contrastive information with the outcome of interest. (3) We further extend our model to accommodate multiple outcome variables, providing a more comprehensive framework for predicting various treatment-specific outcomes simultaneously.

Our proposed CLRM for case-control studies offers several key advantages over existing approaches. First, while current CDR methods are effective in capturing low-dimensional representations unique to the treatment group, they lack the ability to predict (case-specific) outcome variables. 
In contrast, our method directly associates the contrastive information with the treatment group outcome of interest. Second, traditional regression methods do not incorporate case-control structure within the samples, which is a fundamental contribution of our model. Third, our model is designed to naturally quantify uncertainty and accommodate multiple outcomes, further enhancing its applicability across a range of case-control studies. Collectively, these innovations result in a comprehensive and versatile analysis framework for contrastive regression in case-control studies, offering substantial analytic capabilities over existing methodologies.

\section{Contrastive linear regression}
The central focus of contrastive linear regression is to develop a statistical model that captures the variation in the data specific to the `foreground' group (cases), that is not present in the `background' group (controls). In essence, we are interested in the variation that is unique to the case data and the corresponding response variable in foreground group. Any variation in the case-specific response variable that is also present in the control group will lead to less informative associations with the response, we assume, and so we remove shared variation before the associations are modeled. We also assume that the case-control experiment was run properly, so that technical or irrelevant biological effects such as batch or sex were randomized in a balanced way across cases and controls.

We start by defining the contrastive regression model in terms of both the case and control data. From there, we capture shared variation and eliminate it from our model, isolating the case-specific variation. In this method, the response variable is explained through the residual variance remaining in the foreground predictors after controlling for the shared variation.

Consider a set of foreground observations $\{x_i,r_i\}_{i=1}^n\subset \RR^{p}\times \RR$, background observations $\{y_j\}_{j=1}^m\subset\RR^p$, where $x_i, y_j \in \mathbb{R}^p$ and $p$ is typically large, and where $r \in \mathbb{R}$ is the response. We aim to infer the parameter $W$, the loading unique to foreground group, and $\beta$, the regression coefficient. 

While the background observations $\{y_i\}$ are not strictly necessary to fit a regression function for this setting, the motivation for including them is that they enable the characterization of foreground-specific variation, which may be more robust and generalizeable for predicting $r$.
Consider the following model:
\begin{equation}\label{eqn:CLR}
x = Sz_a+Wt+\epsilon_a,~y=Sz_b+\epsilon_b,~r=\beta^\top t+\eta,
\end{equation}
where $z_a,z_b,t\sim N(0,\mathrm{I}_d)$, $\epsilon_a,\epsilon_b\sim N(0,\sigma^2\mathrm{I}_p)$, $\eta\sim N(0,\tau^2)$, $S,W\in\RR^{p\times d}$, $\beta\in\RR^{d}$. Here, $z_a$ and $z_b$ captures the low-dimensional representation of the shared component through loading matrix $S$, $t$ captures the low-dimensional representation of the information unique to the foreground group through loading matrix $W$, and $\epsilon_a$ and $\epsilon_b$ are the residual for $x$ and $y$; $\beta$ is the regression coefficient and $\eta$ is the residual error for $r$.  To simplify the notation, we denote $\theta\coloneqq (S,W,\beta,\sigma^2,\tau^2)$ and $\Theta\coloneqq \RR^{p\times d}\times \RR^{p\times d}\times \RR^d\times \RR_+\times \RR_+$. 

The (log-)likelihood function, which is a measure of the probability of observing the data given the parameters, is derived based on these equations and is maximized to estimate the parameters. Once we have estimated the parameters, we can use them to predict the response for new foreground data or to perform feature selection.

\begin{theorem}\label{thm:likelihood}
Given observations $(X,R)=\{x_i,r_i\}_{i=1}^n$ and $Y=\{y_j\}_{j=1}^m$, the likelihood $p(X,R,Y|\theta)$ is defined as \begin{align}
&p(X,R,Y|S,W,\beta,\sigma^2,\tau^2)=\prod_{i=1}^np(r_i|x_i,S,W,\beta,\sigma^2,\tau)p(x_i|S,W,\sigma^2)\prod_{j=1}^m p(y_j|S,\sigma^2).\label{eqn:likelihood}
\end{align}
The log likelihood $l(\theta)\coloneqq \log p(X,R,Y|\theta)$ is given by 
\begin{align}
    l(\theta) &= -\frac{n}{2}\log \left(\tau^2+\beta^\top A\beta\right)-\frac{1}{2(\tau^2+\beta^\top A\beta)}\sum_{i=1}^n \left(r_i-\beta^\top AW^\top P^{-1}x_i\right)^2\label{eqn:lkhd}\\
    &~~~~-\frac{n}{2}\log |Q|-\frac{1}{2}\sum_{i=1}^nx_i^\top Q^{-1}x_i-\frac{m}{2}\log |P|-\frac{1}{2}\sum_{j=1}^my_j^\top P^{-1}y_j\nonumber,
\end{align}
where $A = (W^\top P^{-1}W+\Id_d)^{-1}\in\RR^{d\times d}$, $P=SS^\top+\sigma^2\Id_p\in\RR^{p\times p}$, and $Q=SS^\top+WW^\top+\sigma^2\Id_p$.
\end{theorem}

Importantly, we can adjust the influence of background data in Equation \eqref{eqn:likelihood} by replacing $p(y_j)$ by $p(y_j)^\alpha$, where $\alpha\geq 0$ is the relative weight of the background data. When $\alpha\to0$, the model evolves into naive linear regression considering only foreground data. Conversely, as $\alpha\to\infty$, model adapts to become a two-step algorithm: first optimizing over $S,\sigma^2$, then over $W,\beta,\sigma^2, \tau$. 

To infer parameters, we use the maximum likelihood estimator (MLE), formulated as $\theta_{ML}=\argmax_{\theta\in\Theta} l(\theta)$. Given the lack of a closed-form maximizer for the log likelihood, we resort to gradient descent. The gradients of the log-likelihood function with respect to each parameter are shown in the following theorem:
\begin{theorem}\label{thm:gradient}
    The gradients of $l$ w.r.t each parameter are given by
\begin{align*}
\frac{\partial l}{\partial S} &=
    -\frac{n\beta\beta^\top AW^\top P^{-2}WA S}{\tau^2+\beta^\top A\beta}+\frac{\beta\beta^\top AW^\top P^{-2}WA S}{2(\tau^2+\beta^\top A\beta)^2}\sum_{i=1}^n \left(r_i-\beta^\top AW^\top P^{-1}x_i\right)^2\\
    &+\frac{2}{(\tau^2+\beta^\top A\beta)} \sum_{i=1}^n \left(r_i-\beta^\top AW^\top P^{-1}x_i\right) \left(\beta^\top W^\top P^{-1}x_i  A W^\top P^{-2} W A S - \beta^\top AW^\top P^{-2} x_i S \right)\\
    &-nQ^{-1}S+\sum_{i=1}^nQ^{-1}x_ix_i^\top Q^{-1}S-mP^{-1}S+\sum_{j=1}^mP^{-1}y_jy_j^\top P^{-1}S,\\
\frac{\partial l}{\partial W} &=
    \frac{n\beta\beta^\top      A(P^{-1}W+W^\top P^{-1})A }{2(\tau^2+\beta^\top A\beta)}+\frac{\beta\beta^\top A(P^{-1}W+W^\top P^{-1})A}{2(\tau^2 +\beta^\top A\beta)^2}\sum_{i=1}^n \left(r_i-\beta^\top AW^\top P^{-1}x_i\right)^2\\
    &-\frac{1}{(\tau^2+\beta^\top A\beta)} \sum_{i=1}^n \left(r_i-\beta^\top AW^\top P^{-1}x_i\right) \left(\beta^\top W^\top P^{-1}x_i A(P^{-1}W+W^\top P^{-1})A   - \beta^\top A P^{-1} x_i \right)\\
    &-nQ^{-1}W+\sum_{i=1}^nQ^{-1}x_ix_i^\top Q^{-1}W,\\
\frac{\partial l}{\partial\beta} &=   
    -\frac{nA\beta}{\tau^2+\beta^\top A\beta}+\frac{A\beta}{(\tau^2+\beta^\top A\beta)^2}     \sum_{i=1}^n\left(r_i-\beta^\top AW^\top P^{-1}x_i\right)^2\\
    &~~~~~+\frac{1}{2(\tau^2+\beta^\top A\beta)}\sum_{i=1}^n(r_i-\beta^\top AW^\top P^{-1}x_i)AW^\top P^{-1}x_i,\\
    \frac{\partial l}{\partial\sigma^2} &=  -\frac{n\beta^\top A^2W^\top        P^{-2}W\beta }{2(\tau^2+\beta^\top A\beta)}+\frac{\beta^\top    A^2W^\top P^{-2}W\beta }{2(\tau^2 +\beta^\top A\beta)^2}\sum_{i=1}^n   \left(r_i-\beta^\top AW^\top P^{-1}x_i\right)^2\\
    &+\frac{1}{\tau^2 +\beta^\top A\beta}\sum_{i=1}^n \left(r_i-\beta^\top AW^\top P^{-1}x_i\right)\left(\beta^\top A^2W^\top P^{-2}W W^\top P^{-1}x_i-\beta^\top AW^\top P^{-2}x_i\right)\\
    &-\frac{n}{2}\tr(Q^{-1})+\frac{1}{2}\sum_{i=1}^n x_i^\top Q^{-2}x_i-\frac{m}{2} \tr(P^{-1})+\frac{1}{2}\sum_{j=1}^m y_j^\top P^{-2}y_j,\\
\frac{\partial l}{\partial\tau^2} &=-\frac{n}{2(\tau^2+\beta^\top A\beta)}+\frac{1}{(\tau^2+\beta^\top A\beta)^2}\sum_{i=1}^n\left(r_i-\beta^\top AW^\top P^{-1}x_i\right)^2.
\end{align*}
\end{theorem}

Once parameters are estimated, we can perform prediction. Given a new input $x_*$ from the foreground group, the goal is to predict the response $r_*$. The predictive distribution is given by the following theorem:
\begin{theorem}\label{thm:predict}
Given parameters $(S,W,\beta,\sigma^2,\tau^2)$, the predictive distribution of $r_*|x_*$ is
\begin{equation}\label{eqn:predict}
 r_*|x_*,S,W,\beta,\sigma^2,\tau^2\sim N\left(\beta^\top AW^\top P^{-1}x_*,\tau^2+\beta^\top A\beta\right).
\end{equation}
As a consequence, the plug-in predictive distribution for $r_*$ is 
\begin{equation}\label{eqn:predict_MLR}
r_*\sim N\left(\beta_{ML}^\top A_{ML}W_{ML}^\top P_{ML}^{-1}x_*,\tau_{ML}^2+\beta_{ML}^\top A_{ML}\beta_{ML}\right).
\end{equation}
Furthermore, the point estimate is $\widehat{r}_*=\beta_{ML}^\top A_{ML}W_{ML}^\top P_{ML}^{-1}x_*$. 
\end{theorem}
That is, we first calculate the MLE, denoted by $\theta_{ML}$, and plug $\theta_{ML}$ into Equation \ref{eqn:predict} to obtain an estimate of $r_*$. Details are given in Algorithm \ref{alg:LL}.

\begin{algorithm}[h]
  \caption{Contrastive linear regression}
  \label{alg:LL}
  \begin{algorithmic}[1]
  \State \textbf{Input:} Training data $\{x_i,r_i\}_{i=1}^n$, $\{y_j\}_{j=1}^m$; new point $x_*$
  \State Find MLE of $S,W,\beta,\sigma^2,\tau$ by maximizing \eqref{eqn:lkhd}.
  \State \textbf{Output:} $r_*\sim N\left(\beta_{ML}^\top A_{ML}W_{ML}^\top P_{ML}^{-1}x_*,\tau_{ML}^2+\beta_{ML}^\top A_{ML}\beta_{ML}\right)$.\\
  \hspace{1.15cm}$\widehat{r}_*=\beta_{ML}^\top A_{ML}W_{ML}^\top P_{ML}^{-1}x_*$.
\end{algorithmic}
\end{algorithm}
In order to demonstrate the flexibility of our method, let us consider a Bayesian approach. By assigning priors to $S,W,\beta,\sigma^2,\tau$, represented by $\Pi$, we can obtain the posterior distribution as shown below:
\begin{equation}\label{eqn:posterior}
\Pi(\theta|X,R,Y)=\Pi(\theta)e^{l(\theta)}
\end{equation}
The posterior predictive distribution for $r_*$ can be computed as follows:
\begin{corollary}\label{clr:prediction}
The posterior predictive distribution for $r_*$ is 
\begin{equation}\label{eqn:predict_posterior}
 p(r_*|x_*,S,X,R,Y)= \int_{\Theta}N\left(r_*;\beta^\top AW^\top P^{-1}x_*,\tau^2+\beta^\top A\beta\right)\Pi(\theta|X,R,Y)\mathrm{d}\theta.
\end{equation}
\end{corollary}

Correspondingly, a Bayesian variant of our algorithm can be derived, using the posterior for parameter estimation, as described in Algorithm \ref{alg:LL_Bayesian}.
\begin{algorithm}[h]
  \caption{Bayesian contrastive linear regression}
  \label{alg:LL_Bayesian}
  \begin{algorithmic}[1]
  \State \textbf{Input:} Training data $\{x_i,r_i\}_{i=1}^n$, $\{y_j\}_{j=1}^m$; new point $x_*$, number of posterior samples $T$. 
  \For{t=1:T}
    \State Sample $\theta_t=(S_t,W_t,\beta_t,\sigma_t^2,\tau_t)$ from the posterior given by \eqref{eqn:posterior}.
    \State Sample $r_t$ from $N\left(\frac{\tau\beta^\top B^{-1}(W^\top x_*-W^\top SA^{-1}S^\top x_*)}{\sigma^2(\tau-\beta^\top B^{-1}\beta)},\frac{\tau^2}{\tau-\beta^\top B^{-1}\beta}\right)$.
   \EndFor
\end{algorithmic}
\end{algorithm}

\section{Experiments}

Next, we show how contrastive regression behaves when applied to diverse case-control study data where the cases include a response variable. First, we consider simulation studies to validate our approach using predictive performance, accuracy, and consistency metrics. Next, we apply contrastive regression to an image dataset to illustrate its behavior in a visual context. Then we apply contrastive regression to a single-cell RNA-sequencing study of nasal samples with and without nasal polyps. Finally, we apply contrastive regression to a single nucleus RNA-sequencing brain dataset from people with and without autism.

\subsection{Simulation studies}
\subsubsection{Predictive performance and time complexity}

\begin{figure}
    \centering
    \includegraphics[width=\textwidth]{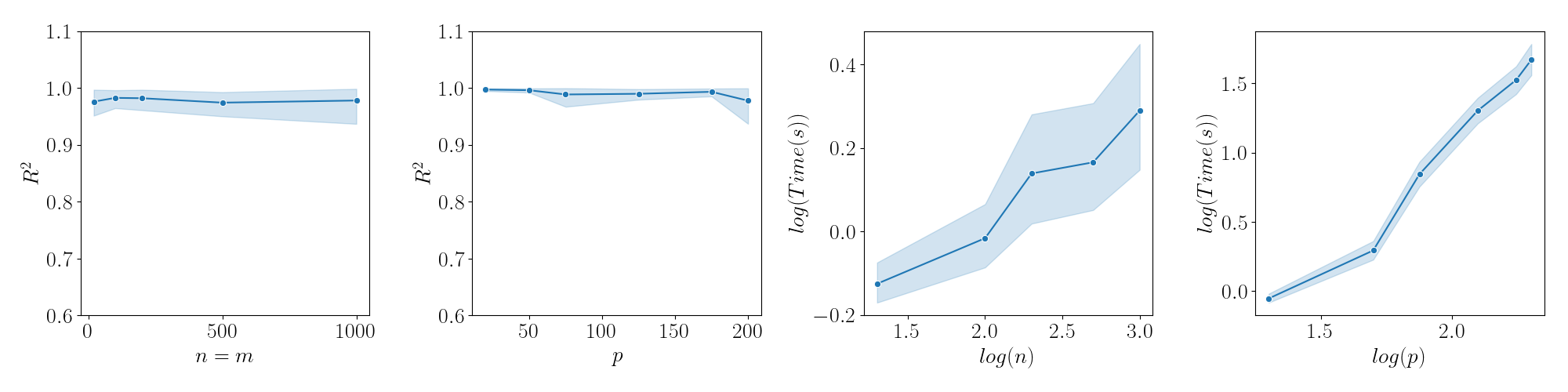}
    \caption{\textbf{Simulation study for contrastive regression: $R^2$ and time.}
    We vary the number of samples on the x-axis. The left two panels show the $R^2$ between predictions and ground truth for a simulated dataset. The right two panels show the time required to fit the model. The ribbons in both capture 95\% confidence intervals.
    }
    \label{fig:sample_size_p}
\end{figure}

We evaluate the model performance and time complexity with different combinations of sample size and feature dimension. First, we generated data from the model with a fixed feature dimension ($p$ = 2) and a varying sample size, ranging from 20 to 1,000. We fit the model with tolerance $0.0001$, compute $R^2$ between predictions and truth, and measured time in seconds. As sample size increased, we observed that $R^2$ remained near one. Second, we generated the data from the model with a fixed sample size ($n = m = 200$) and a varying feature dimension, ranging from $20$ to $200$. We found that $R^2$ was not affected by the feature dimension. In both scenarios, it took less than $100$ seconds to fit the model, and the time complexity is $\mathcal{O}((n+m)p^2)$. (\autoref{fig:sample_size_p})

\subsubsection{Estimation accuracy}

\begin{figure}[!ht]
    \centering
    \includegraphics[width=\textwidth]{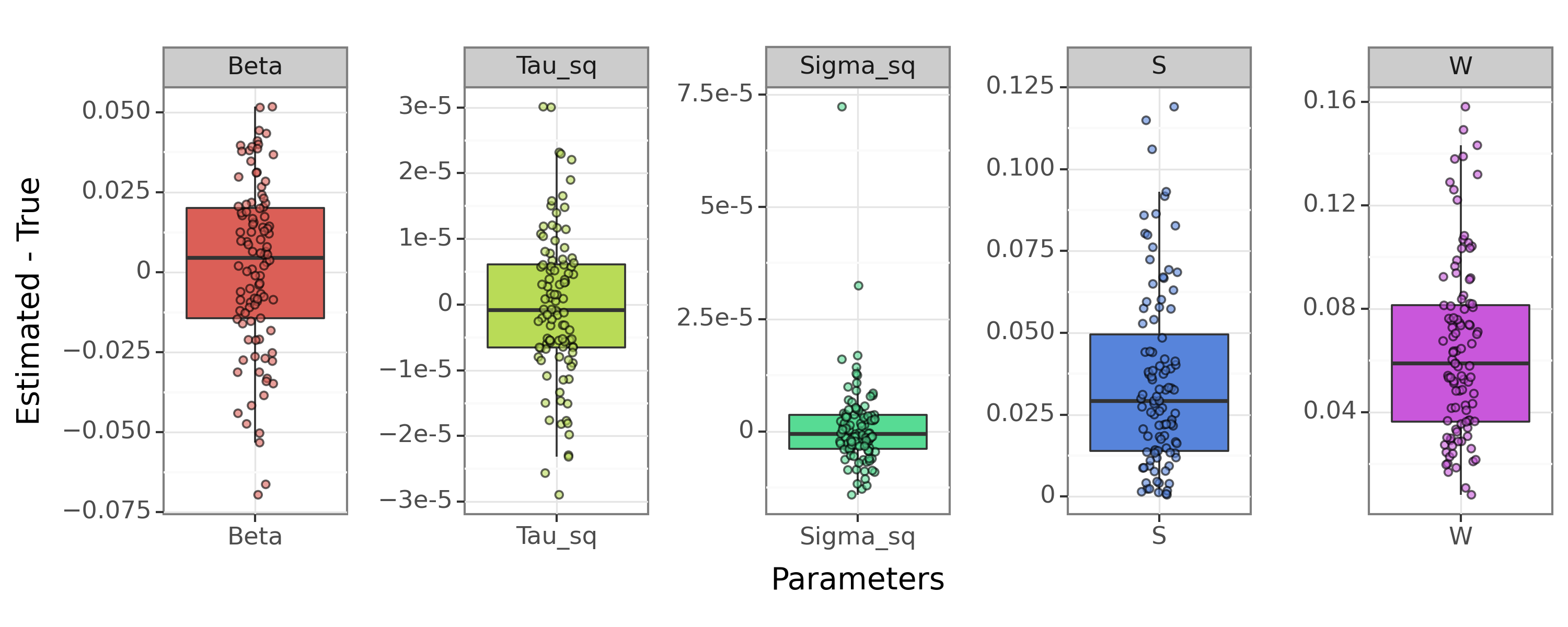}
    \vspace{-10mm}
    \caption{\textbf{Estimation accuracy for contrastive regression.}
    The difference between estimated and true value for each of the parameters. The first three panels are the total difference between estimated and true values. The y-axis for $S$ and $W$ is defined as $\|\hat{S}\hat{S}^T - SS^T\|/\|SS^T\|$ and $\|\hat{W}\hat{W}^T - WW^T\|/\|WW^T\|$, respectively.
    }
    \label{fig:est_error}
\end{figure}

To quantify the estimation error for the parameters, we 
generate data from the contrastive regression model with a fixed sample size ($n = m = 500$) and a fixed feature dimension ($p = 2$). We fit the model, repeat the same procedure $100$ times, and compute the estimation error for each parameter. Here, the estimation error is defined as the difference between the parameter estimate and the ground truth. We quantify the estimation error for $\beta$ by taking the difference of absolute value of the point estimate and the ground truth. For the estimation of $\tau^2$ and $\sigma^2$, the estimation error is quantified by taking the difference of point estimate and ground truth, and they  are less than $7.5\times 10^{-5}$. Due to nonidentifiablity issues, we define the error for $S$ and $W$ as $\|\hat{S}\hat{S}^T - SS^T\|/\|SS^T\|$ and $\|\hat{W}\hat{W}^T - WW^T\|/\|WW^T\|$, respectively. The estimation errors are well controlled in our inference (\autoref{fig:est_error}).

\subsubsection{Consistency of inference as sample size increases}

\begin{figure}[!ht]
    \centering
    \includegraphics[width=\textwidth]{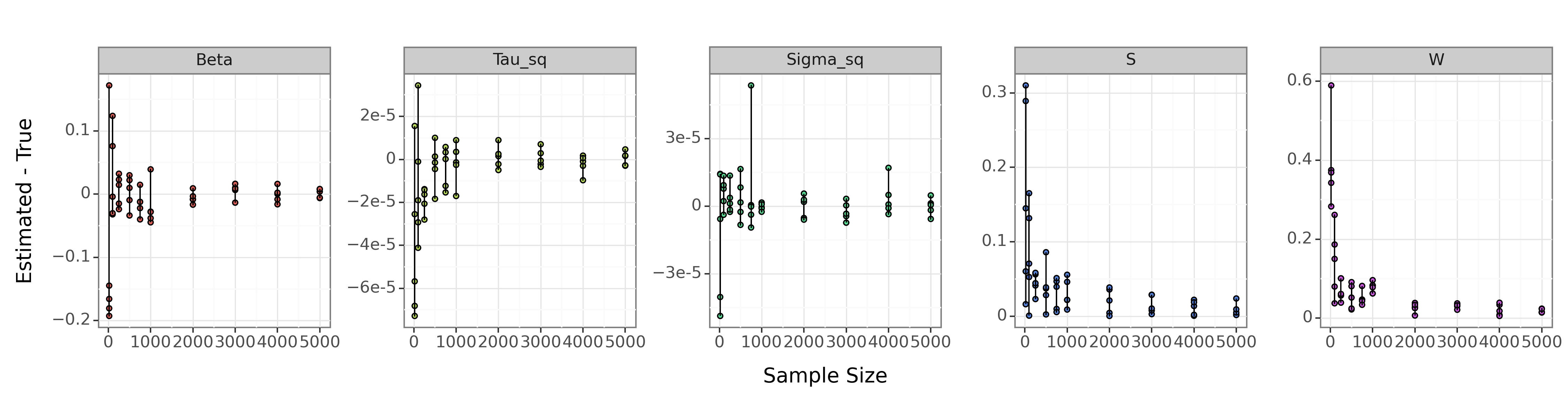}
     \vspace{-10mm}
    \caption{\textbf{Consistency of inference as sample size increases.}
    The difference between the parameter estimate and the true value (y-axis) for each parameter contracts around 0 when sample size (x-axis) increases. 
    }
    \label{fig:est_consistency}
\end{figure}

To demonstrate the consistency of the estimators, we again simulated data from the contrastive regression model; we increased the sample size ($n=m$) from 20 to 5000 with a fixed feature dimension ($p = 2$). We find that the estimates are closer to the truth as the sample size increases for all parameters (\autoref{fig:est_consistency}). 

\subsubsection{Contrastive regression behavior on the corrupted lines dataset}
\begin{figure}
    \centering
    \includegraphics[width=\textwidth]{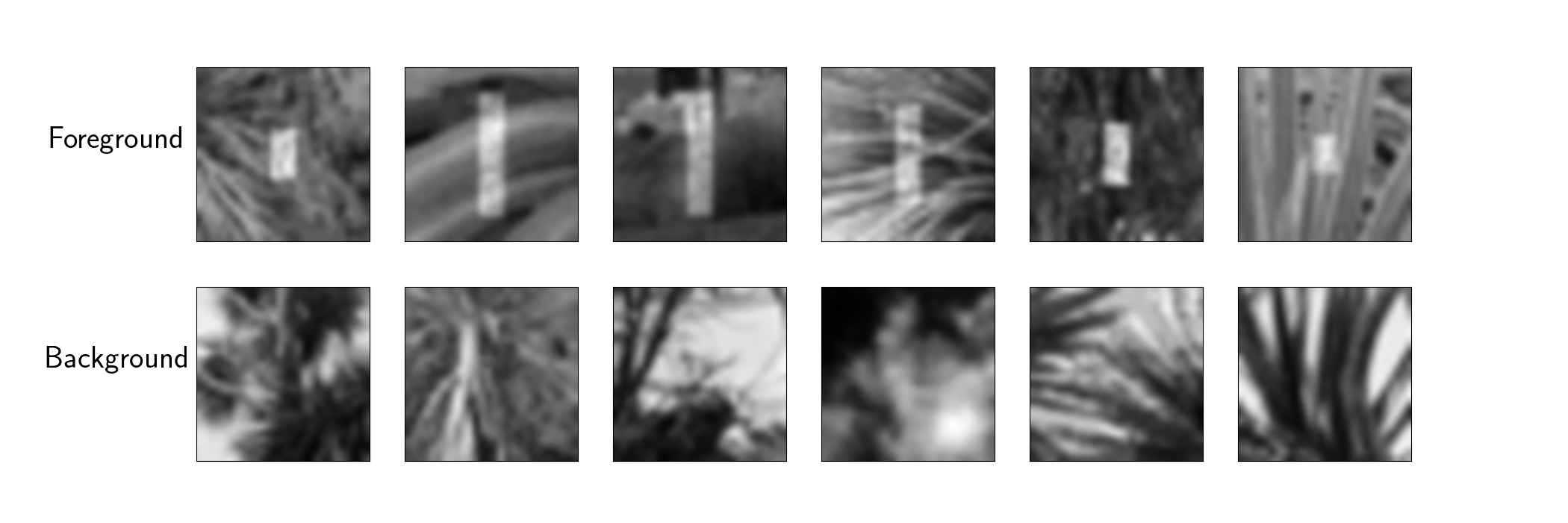}
    \caption{\textbf{Contrastive regression applied to the corrupted lines dataset.}
    The top row shows example foreground samples, and the bottom row shows example background samples. We treat the height of the line in each foreground image as the foreground-specific response variable $r$.
    }
    \label{fig:corrupted_lines_example}
\end{figure}

To demonstrate the behavior of our model in a visual context, we used a dataset of images that we call the \emph{corrupted lines} dataset. This dataset, inspired by the corrupted MNIST dataset~\citep{abid2018exploring}, consists of natural images of plants in the background (\autoref{fig:corrupted_lines_example}). The foreground data is also made up of natural images of plants, but these images have been corrupted by a vertical line. The line in each image has a varying height. We treat the image pixel values as the foreground and background data $X$ and $Y$, and the line height as the response $r$.

\begin{figure}[!h]
    \centering
    \includegraphics[width=\textwidth]{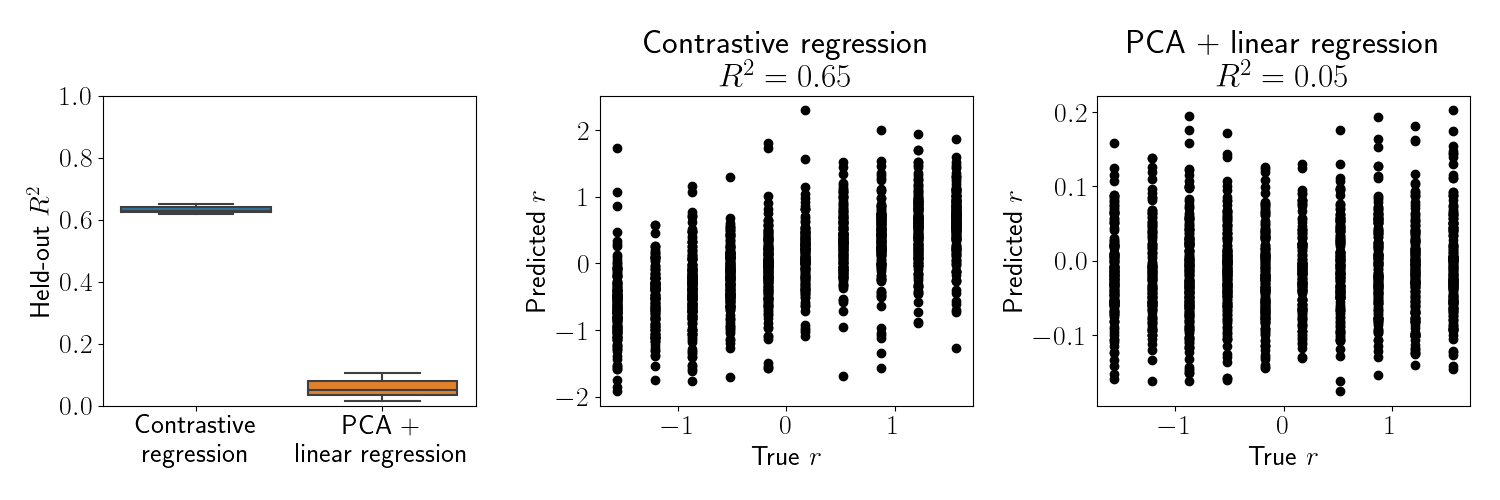}
    \caption{\textbf{Corrupted lines experiment.}
    \emph{Left:} Goodness-of-fit $R^2$ between foreground-specific responses (line heights) and predictions across three train/test splits. 
    \emph{Middle:} True responses plotted against predictions from the contrastive regression model for one of the train/test splits (test set $R^2 = 0.65$).
    \emph{Right:} Same as the middle panel, but for PCA plus linear regression model (test set $R^2 = 0.05$).
    }
    \label{fig:corrupted_lines_preds}
\end{figure}

We fit our contrastive regression model on the corrupted lines dataset using a latent dimension of $d=2$ and then conducted a prediction experiment. Specifically, we fit the model using 67\% of the samples selected at random and held out the remaining 33\% of the samples for testing. We then computed predictions for the held-out responses, and we measured the goodness-of-fit $R^2$ value between the predicted line lengths and true line lengths. For comparison, we also fit PCA plus linear regression with $d=2$ latent components. The contrastive regression model had better predictive performance than PCA plus linear regression, with $R^2 = 0.65$ for contrastive regression versus $R^2 = 0.05$ for PCA plus linear regression (\autoref{fig:corrupted_lines_preds}).

\subsection{Application to cellular differentiation trajectories}
We applied our contrastive regression model to single cell RNA-seq (scRNA-seq) data, and in particular to cells undergoing differentiation. We considered a study of cellular differentiation in the nasal epithelium comparing two different disease states -- chronic rhinosinusitis with nasal polyps and chronic rhinosinusitis without nasal polyps. Prior work sought to identify the underlying differences between these two phenotypes and found that, in patients with nasal polyps, cellular differentiation of epithelial cells from basal to secretory cells was dysregulated, leading to stunted differentiation progress~\citep{ordovas2018allergic}. To quantify this cellular differentiation trajectory, the authors used a pseudotime algorithm, ordering cells in a representation of differentiation space based on their transcriptomes. 

\begin{figure}[!h]
    \centering
    \includegraphics[width=\textwidth]{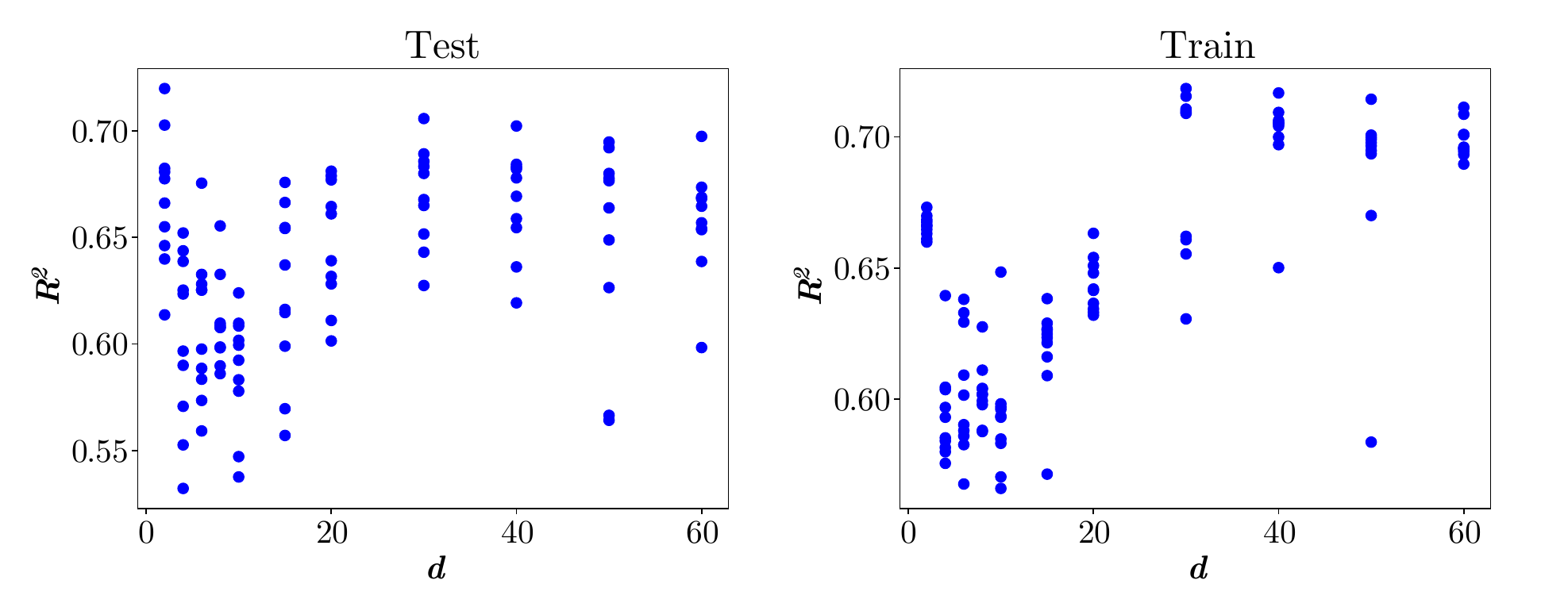}
    \caption{\textbf{Cross validation for selection of $d$.} Results of 10-fold cross-validation on nasal polyps cellular differentiation comparison. $R^2$ calculated between published differentiation trajectory order and predicted ordering of cells in training splits (right) and test splits (left) for selected values of $d$.
    }
    \label{fig:cv}
\end{figure}

To apply contrastive regression to these data, we let $X$, the foreground, be the matrix of normalized scRNA-seq counts from epithelial cells derived from non-polyp samples, and let $Y$, the background, be a matrix of normalized scRNA-seq counts from epithelial cells derived from polyp samples. We define the response variable $r$ to be a cell's location on the developmental trajectory. We chose the dimension $d$ using 10-fold cross-validation (Figure \ref{fig:cv}) and ran the model on the top $500$ variable genes. Here, the optimal $d$ value was $3$, resulting in an $R^2$ of $0.69$ (Figure \ref{fig:nopolyprsquared}). To determine the latent dimension of interest in the non-polyp specific, or foreground, matrix $W$, we selected the dimension with the largest absolute value of $\beta$, and we examined the genes with the highest $W$ loadings. 

\begin{figure}[!h]
    \centering
    \includegraphics[width=\textwidth]{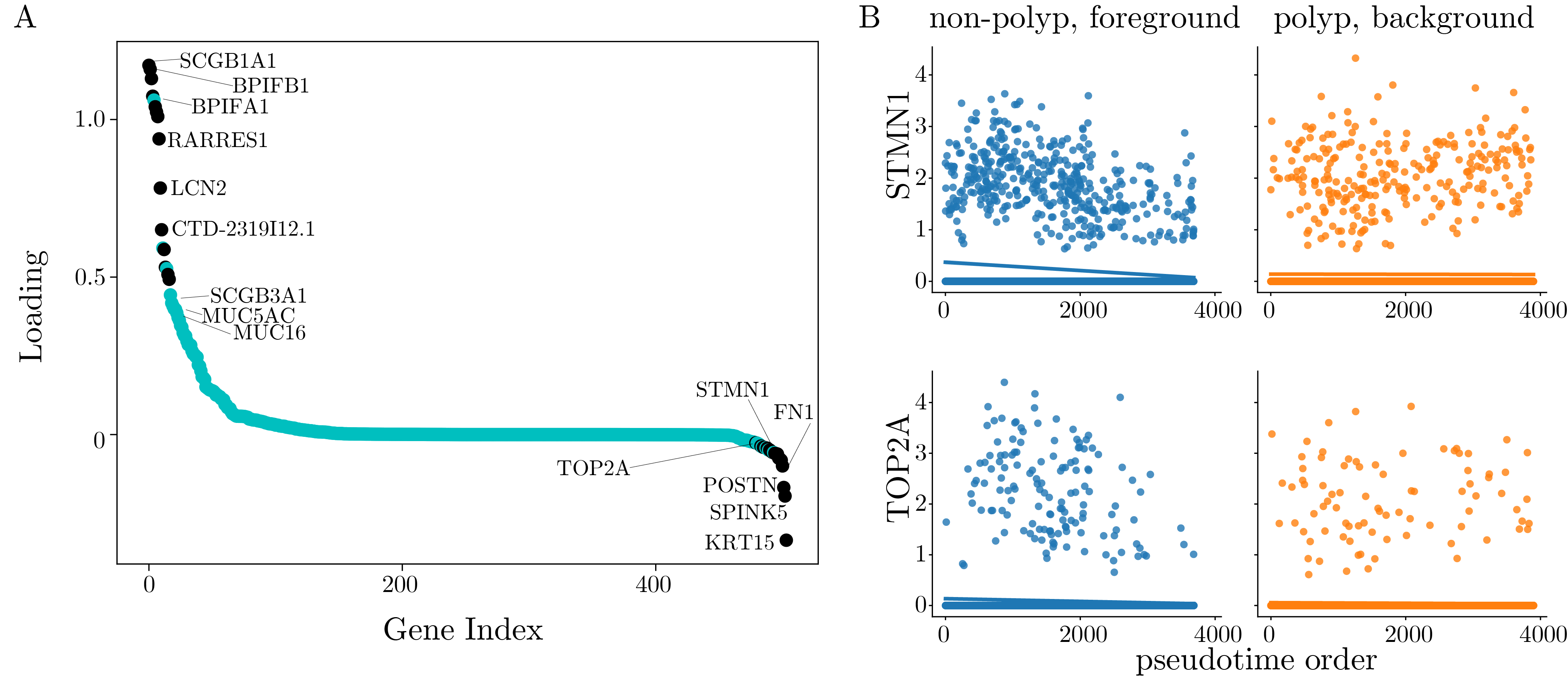}
    \caption{\textbf{Nasal polyp genes}. (A) Top 500 variable genes ordered by contrastive regression loadings from the latent component with the largest absolute value of $\beta$. Black points indicate genes identified using previous method as differentially correlated between polyp and non-polyp differentiation trajectories, and blue points are genes not identified by this method. Select genes are labeled. (B) $\log_{10}$-normalized gene expression values of two cell-cycle related genes identified via contrastive regression but not prior methods. First column (blue) shows per-cell expression of each gene in cells from non-polyp (foreground) patients ordered by the pseudotime trajectory order (\textit{r}). Second column (orange) shows this gene expression in cells derived from donors with nasal polyps ordered along the trajectory and showing little correlation with trajectory location.
    }
    \label{fig:np}
\end{figure}

In the original study, genes related to the changes in differentiation were selected based on differences in the Spearman correlation of their expression values with pseudotime trajectory between non-polyp cells and polyp cells. Genes with high loadings values using contrastive regression agreed largely with genes identified via the Spearman correlation. Contrastive regression identified a few additional genes, including some involved in mitosis (\emph{STMN1, TOP2A}), as well as genes related to epithelial barrier immunity (\emph{SCGB3A1, BPIFA1, MUC5AC, MUC16}; Figure \ref{fig:np} A). These immunity-related genes complement genes identified by the original analysis method such as \emph{SCGB1A1}, a secretoglobin which, like \emph{SCGB3A1}, is involved in immune response in upper airways in differentiated secretory cells~\citep{mootz2022secretoglobins}. In addition, the gene \emph{BPIFA1}, newly identified by our model, taken with the gene \emph{BPIFB1}, which was also previously identified, are related to mucus secretion expressed in differentiated epithelial cells in the upper airway especially during infection~\citep{de2017association}. Contrastive regression also retained genes considered by the authors of the original study to be relevant, including extracellular matrix components (\emph{POSTN, FN1}), and mediators of Wnt and Notch signaling (\emph{DLK2, DLL1, JAG2}).  Contrastive regression expanded the list of differential genes as compared to prior work.

Since it is possible to assign pseudotime values to any cell, even those not actively differentiating, the original study was able to give pseudotime values to the polyp cells as well as the non-polyp cells. While contrastive regression did not include the pseudotime values of polyp cells, because they were considered background, we used these data to have ground truth for a response for background data to confirm the signal from the shared components is associated with the masked response. We compared the relationship between gene expression value and pseudotime for these top genes in the polyp compared to the non-polyp cells, and we found that many of these genes either increased or decreased in expression in the non-polyp cells over pseudotime but not in the polyp cells (Figure \ref{fig:np}B, Figure \ref{fig:supp_polyp_genes}). Contrastive regression was therefore able to identify genes with differential correlation with pseudotime between these two groups without direct knowledge of the pseudotime ordering in the background (polyp) group. 

\subsection{Application to autism single-nucleus RNA-seq data}

To further study the behavior of contrastive regression on case-control genomic data, we next applied our model to a single-nucleus RNA sequencing (snRNA-seq) dataset from prefrontal cortex (PFC) postmortem tissue samples from patients with autism spectrum disorder (ASD) and controls~\citep{Velmeshev2019autism}. Gene expression profiles, cell-type annotations, and clinical severity of ASD are associated with each patient. In this example, we use sample-level phenotype, which is clinical severity of ASD, as the response variable. We focus on PFC samples for downstream analysis.

\begin{figure}[!h]
    \centering
    \includegraphics[width=\textwidth]{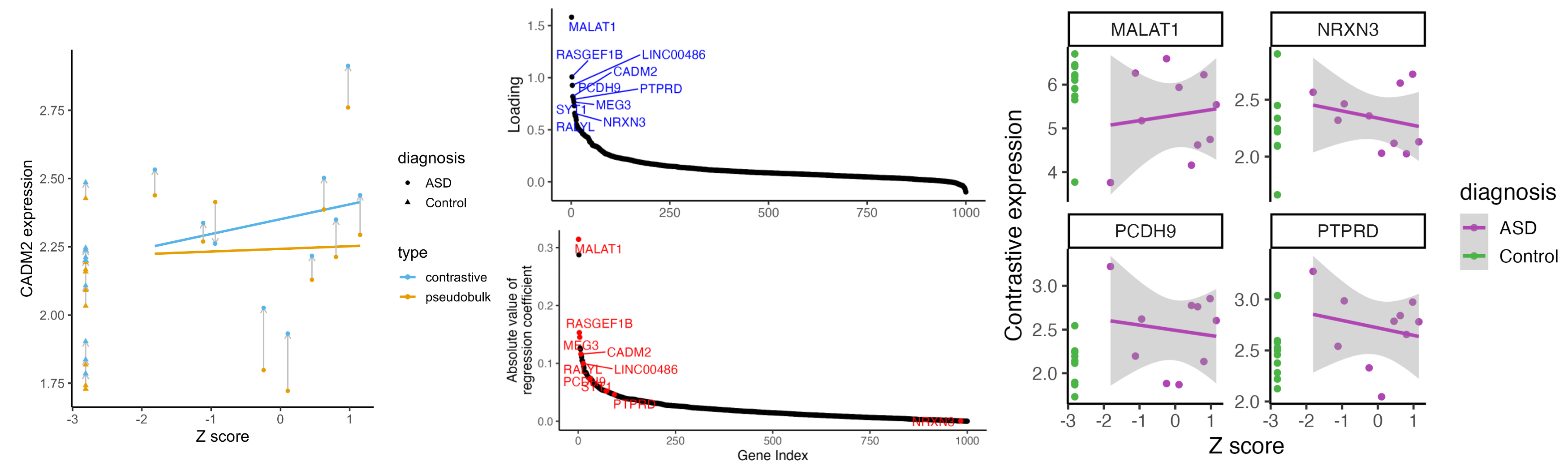}
    \caption{\textbf{Contrastive regression identifies autism-related genes in L2/3 excitatory neurons.}
    \emph{Left: Gene expression of \emph{CADM2} plotted against Z-score of ADI-R social interaction subscore, with shape representing clinical diagnosis, and color indicating pseudobulk or contrastive expression levels} . 
    \emph{Middle - upper panel: Top 1000 highly variable genes were ordered by contrastive loadings from a latent component with largest absolute value of $\beta$, and top ten genes were highlighted in blue} .
    \emph{Middle - lower panel: Top 1K highly variable genes were ordered by absolute value of regression coefficient, and top $10$ genes identified using contrastive regression were highlighted in red} .
    \emph{Right: Contrastive expression of \emph{MALAT1}, \emph{NRXN3}, \emph{PCDH9} and \emph{PTPRD} against Z-score of ADI-R social interaction subscore. 
    } .
    }
    \label{fig:asd_genes}
\end{figure}

We fit the contrastive regression to the snRNA-seq data. We identified genes associated with the Autism Diagnostic Interview-Revised (ADI-R) score A (social interactions). Here we filtered out autism patients without social interaction score. For each cell type, we first pooled cells from each sample to form a sample-level pseudobulk expression matrix. Genes that are expressed in less than 1\% cells were further excluded from the analysis. The top 1000 highly variable genes (HVGs) were identified using the \textit{getTopHVGs()} function from the \textit{scran} R package~\citep{Lun2016scran}. We subset the data to the most variable 1000 genes for downstream analysis. Then, we let $X$, the foreground data, be the normalized pseudobulk expression of $10$ autism patients, and we let $Y$, the background data, be the normalized pseudobulk expression of $10$ controls. Z-score transformation of ADI-R social interaction score was used as the response variable $r$. We fit the contrastive regression model with $d = 8$ latent components. To determine the latent component of interest in the autism-specific matrix $W$, we chose the dimension with the largest absolute value of $\beta$, and we examined the genes with the highest $W$ loadings (\autoref{fig:asd_genes}).

To better illustrate how contrastive regression works, we compared the expression profile before and after applying our method (\autoref{fig:asd_genes}). Contrastive expression values are the residuals after eliminating the shared component ($Sz$) by minimizing the reconstruction error. We observed a more extreme positive association between \emph{CADM2} expression and social interaction score in ASD patients in the contrastive expression results as compared to standard pseudobulk expression analyses (\autoref{fig:asd_genes}).

In the layer 2/3 excitatory neurons (L2/3) cell type, we find that the top-ranking genes identified by contrastive regression are all ASD-related (\autoref{fig:asd_genes}; \autoref{fig:supp_asd_genes}A-B). Despite its crucial role as a prognostic marker in several cancers~\citep{Dong2014malat1ost,Lu2016malat1cervical, Meseure2016malat1breast}, \emph{MALAT1} (also known as \emph{NEAT2}), a long noncoding RNA (lncRNA), has been previously reported as a differential signature comparing ASD individuals and controls~\citep{Voineagu2011malat1diff, Parikshak2013malat1diff} (\autoref{fig:asd_genes}, right). \emph{MALAT1} has been shown to be highly expressed in neurons and to regulate synapse formation~\citep{Bernard2010malat1func}. \emph{NRXN3} is one of the neurexin family genes. The neurexin gene family encodes cell adhesion molecules and plays important roles in synaptic development and neurotransmitter release~\citep{Missler2003nrxn3} (\autoref{fig:asd_genes}, right). Rare deletions within the \emph{NRXN3} locus have been identified in ASD patients~\citep{Vaags2012nrxn3}. \emph{NRXN3} has also been reported as one of the ASD risk genes with high confidence in the Simons Foundation Autism Research Initiative (SFARI) database~\citep{BanerjeeBasu2010SFARI}. \emph{PCDH9}, one of the strong candidate genes associated with ASD in the SFARI database, is involved in cell adhesion, and contributes to the formation of functional neuronal circuits~\citep{Mancini2020pcdh9} (\autoref{fig:asd_genes}, right). Prior work has identified rare mutations in \emph{PCDH9} gene~\citep{Marshall2008pcdh9}.  Other research localized \emph{PCDH9} in the neural synapse, and showed that \emph{PCDH9} knockout mice result in defects in the nervous system and abnormal emotional behaviors resembling ASD-like phenotypes~\citep{Uemura2022pcdh9}. \emph{PTPRD} is a receptor protein tyrosine phosphatase highly expressed in the brain, and has been associated with various neural disorders including ASD~\citep{Tomita2020ptprd} (\autoref{fig:asd_genes}, right). Prior work hypothesized that \emph{PTPRD} regulates neurogenesis by modulating the RTK-MEK-ERK signaling pathway, and that the loss of \emph{PTPRD} function can lead to cognitive alterations~\citep{Tomita2020ptprd}. 

To confirm these results, we carried out a sensitivity analysis where we excluded three outliers (one control and two cases); the results remain largely unchanged (\autoref{fig:supp_sensitivity}). We set the ADI-R score to 0 for the controls and recomputed the z-score transformation across cases and controls. Then we applied linear regression with the normalized gene expression profile as the response variable and the z-score as explanatory variable. We ranked the genes by the absolute value of regression coeffcient and found that the majority of top selected genes using contrastive regression remained at the top when linear regression was applied (\autoref{fig:asd_genes}). However, traditional linear regression fails to identify one of the most important ASD risk gene, \emph{NRXN3} (\autoref{fig:asd_genes}). This demonstrates that contrastive linear regression can retrieve biologically important signals after removing shared variation between cases and controls.   

Through visualizing the shared latent factors for each sample, we observed that the shared variation is able to partially separate ASD samples and controls. By plotting out the pairwise latent dimensions, we find that the controls cluster more tightly compared to the ASD cases, indicating that the ASD samples are more heterogeneous (\autoref{fig:supp_asd_genes}C).

\begin{figure}[!h]
    \centering
    \includegraphics[width=\textwidth]{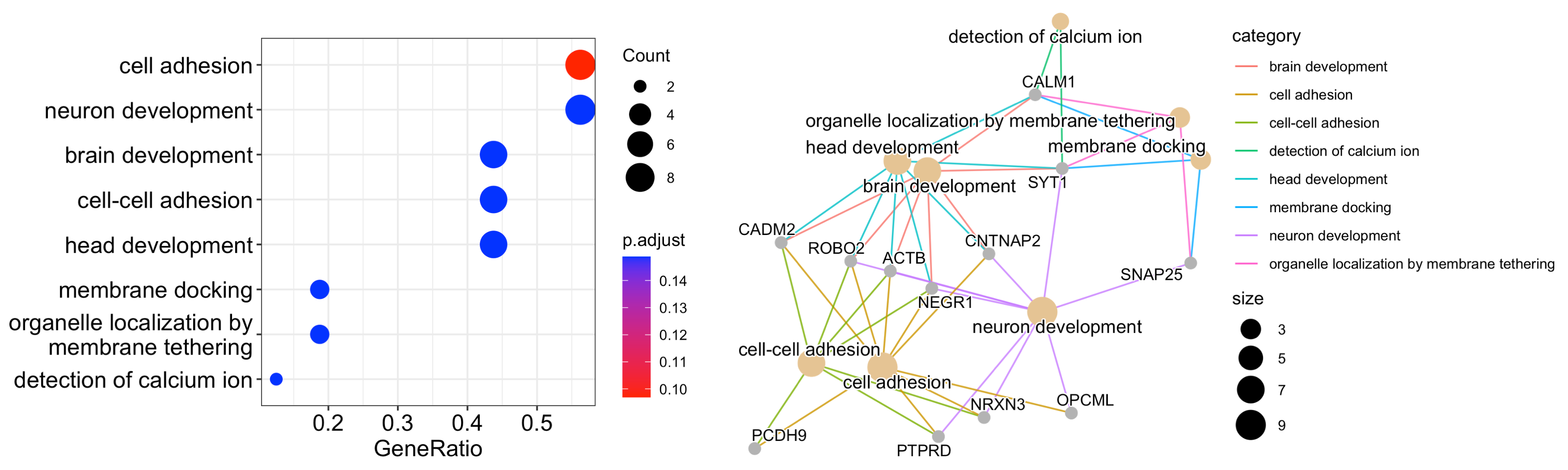}
    \caption{\textbf{Enriched biological process GO terms of top 20 genes in L2/3 excitatory neurons.}
    \emph{Left: Eight biological processes related GO terms were substantially overrepresented among the top 20 genes identified by contrastive regression, with adjusted p-value $\leq$ 0.15} . 
    \emph{Right: Linkage between genes and GO terms in the network plot, with line color indicating eight GO terms and circle size representing the number of top genes that were found for each GO term} .
    }
    \label{fig:asd_pathway}
\end{figure}

To better summarize the underlying biological process, we performed a GO enrichment analysis~\citep{goAshburner2000-gt, goGene_Ontology_Consortium2023-xg} and a KEGG pathway analysis~\citep{keggKanehisa2000-yb, keggKanehisa2019-vn, keggKanehisa2023-cm} on the ranked genes. Cell adhesion was the most significant term with adjusted p-value $< 0.1$ in both analyses (\autoref{fig:asd_pathway}). Cell adhesion molecules are critical for neuronal function, especially synapse formation. Previous studies have shown that multiple genes, including genes in the neurexin and protocadherin families, encode synaptic cell adhesion molecules and variants in these genes are associated with an increased risk of ASD. These results suggest that contrastive regression can model the variability in clinical responses in the foreground data by removing irrelevant variation shared with the background data that masks the variation unique to the cases.

\section{Discussion}

In this work, we developed a contrastive linear regression model that controls for shared variation in the predictors for cases and controls before performing regression on the case-specific responses. We validated our approach on two case-control data sets: One considered single-cell gene expression in nasal epithelium comparing chronic rhinosinusitis with nasal polyps (controls) and chronic rhinosinusitis without nasal polyps (cases) with pseudotime information, and another used gene expression data from the brain in cases with autism including scores capturing the severity of the disease and in controls without these scores.

A number of possible extensions could improve our method. In many case-control experiments, it is possible that the controls also have responses. In this case, we can include contrastive regression for the controls after accounting for shared variation between cases and controls. We can also consider extending either the regression or the contrastive latent variable model to a nonlinear model. The results for both of the applications show possible nonlinear associations in the contrastive regression, suggesting that Gaussian process regression may be more appropriate than linear regression. Finally, it may be the case that the responses are count based or binary. In this situation, we can consider extending these models to generalized linear models to allow better representation of these data likelihoods.



\begin{appendix}
\section{Code availability}
Code for the model and experiments is available at \texttt{\url{https://github.com/byzhang23/contrastive-regression}}.

\section{Data availability}

\subsection{Corrupted lines dataset}
Grass images were obtained from ImageNet~\citep{deng2009imagenet}.

\subsection{Nasal Polyp scRNA-seq dataset} Raw count data, pseudotime values, and metadata were downloaded from the original article~\citep{ordovas2018allergic} at \texttt{\url{https://singlecell.broadinstitute.org/single_cell/study/SCP253}}
\subsection{Autism snRNA-seq dataset} Sample-level metadata was downloaded from Supplementary Material - Data S1 in the original article~\citep{Velmeshev2019autism}. Raw count matrix and cell-level metadata were downloaded from \texttt{\url{https://autism.cells.ucsc.edu}}.
\section{Proof of Theorem \ref{thm:likelihood}}
First observe that 
\begin{align}
&p(X,R,Y|S,W,\beta,\sigma^2,\tau^2)=\prod_{i=1}^np(r_i|x_i,S,W,\beta,\sigma^2,\tau)p(x_i|S,W,\sigma^2)\prod_{j=1}^m p(y_j|S,\sigma^2)\nonumber\\
    &= \prod_{i=1}^nN(r_i;\beta^\top AW^\top P^{-1}x_i,\tau^2+\beta^\top A\beta)N(x_i;0,Q)\prod_{j=1}^m N(y_j;0,P)  ,
\end{align}
where $P=SS^\top+\sigma^2\Id_p\in\RR^{p\times p}$, $Q=SS^\top+WW^\top+\sigma^2\Id_p$, and $A = (W^\top P^{-1}W+\Id_d)^{-1}\in\RR^{d\times d}$. 
It's clear that $y\sim N(0,SS^\top+\sigma^2\mathrm{I}_p)$ and $x\sim N(0,SS^\top+WW^\top+\sigma^2\mathrm{I}_p)$, our goal is to derive $p(r|x)$:
$$p(r|x)=\int_{\RR^d} p(r|x,t)p(t|x)dt=\int_{\RR^d} p(r|t)p(t|x)dt.$$

Observe that $p(t|x)=\frac{p(x|t)p(t)}{p(x)}$ and we already have
\begin{align*}
    x|t\sim N(Wt,P),~x\sim N(0,Q),~t\sim N(0,\Id_d).
\end{align*}
where $P=SS^\top+\sigma^2\Id_p$ and $Q = S S^\top+WW^\top+\sigma^2\mathrm{I}_p$. \begin{align*}
p(t|x) &\propto  e^{-\frac{1}{2}(x-Wt)^\top P^{-1}(x-Wt)}e^{-\frac{1}{2}t^\top t}e^{\frac{1}{2}x^\top Q^{-1}x}\\
& = \propto e^{-\frac{1}{2}t^\top \left(W^\top P^{-1}W+\Id_d\right)t+x^\top P^{-1}Wt}
\end{align*}

As a result, 
\begin{align*}
   t|x&\sim N((W^\top P^{-1}W+\Id_d)^{-1}W^\top P^{-1}x,(W^\top P^{-1}W+\Id_d)^{-1})\\
   &=N(AW^\top P^{-1}x,A),\end{align*}
where $A\coloneqq (W^\top P^{-1}W+\Id_d)^{-1}$.

Now we can calculate $p(r|x)$. Observe that $r|x$ is Gaussian, it suffices to find the mean and variance of $r$. 
\begin{align*}
    \EE[r|x] &= \EE[\EE[r|x,t]]=\EE[\beta^\top t|x] = \beta^\top AW^\top P^{-1}x\\
    \var[r|x]& = \EE[\var[r|t]]+\var[\EE[r|t]|x]= \tau^2+\beta^\top A\beta
\end{align*}
To conclude
\begin{equation}\label{eqn:r|x}
    r|x\sim N\left(\beta^\top AW^\top P^{-1}x,\tau^2+\beta^\top A\beta\right)
\end{equation}
\begin{align*}
l(S,W,\beta,\sigma^2,\tau)&\coloneqq  \log L(S,W,\beta,\sigma^2,\tau)\\
&=\sum_{i=1}^n\log p(r_i|x_i,S,W,\beta,\sigma^2,\tau)+\sum_{i=1}^n\log p(x_i|S,W,\sigma^2)+\sum_{j=1}^m \log p(y_j|S,\sigma^2).
\end{align*}
By previous notation, we can simplify $l$:
\begin{align}
    &l(S,W,\beta,\sigma^2,\tau) = \sum_{i=1}^n \left(-\frac{1}{2}\log \left(\tau^2+\beta^\top A\beta\right)-\frac{1}{2}\frac{\left(r_i-\beta^\top AW^\top P^{-1}x_i\right)^2}{\tau^2+\beta^\top A\beta}\right)\nonumber\\
    &+\sum_{i=1}^n \left(-\frac{1}{2}\log|Q|-\frac{1}{2}x_i^\top Q^{-1}x_i\right)+\sum_{j=1}^m \left(-\frac{1}{2}\log|P|-\frac{1}{2}y_j^\top P^{-1}y_j\right)\nonumber\\
    & = -\frac{n}{2}\log \left(\tau^2+\beta^\top A\beta\right)-\frac{1}{2(\tau^2+\beta^\top A\beta)}\sum_{i=1}^n \left(r_i-\beta^\top AW^\top P^{-1}x_i\right)^2\nonumber\\
    &~~~~-\frac{n}{2}\log |Q|-\frac{1}{2}\sum_{i=1}^nx_i^\top Q^{-1}x_i-\frac{m}{2}\log |P|-\frac{1}{2}\sum_{j=1}^my_j^\top P^{-1}y_j\nonumber.
\end{align}

\section{Proof of Theorem \ref{thm:gradient}}
Recall that the log-likelihood $l(\theta)$ is given by 
\begin{align}
    l(\theta) &= -\frac{n}{2}\log \left(\tau^2+\beta^\top A\beta\right)-\frac{1}{2(\tau^2+\beta^\top A\beta)}\sum_{i=1}^n \left(r_i-\beta^\top AW^\top P^{-1}x_i\right)^2\label{eqn:lkhd}\\
    &~~~~-\frac{n}{2}\log |Q|-\frac{1}{2}\sum_{i=1}^nx_i^\top Q^{-1}x_i-\frac{m}{2}\log |P|-\frac{1}{2}\sum_{j=1}^my_j^\top P^{-1}y_j\nonumber,
\end{align}
where $A = (W^\top P^{-1}W+\Id_d)^{-1}\in\RR^{d\times d}$, $P=SS^\top+\sigma^2\Id_p\in\RR^{p\times p}$, and $Q=SS^\top+WW^\top+\sigma^2\Id_p$.

Now we can calculate the gradient of $l$ w.r.t. $(S,W,\beta,\sigma^2,\tau^2)$:

For $\frac{\partial l}{\partial S}$, we need to following building block:
\begin{align*}
\frac{\partial P}{\partial S } = 2S,~\frac{\partial A}{\partial P } = AW^\top P^{-2}WA,~\frac{\partial A}{\partial S} =\frac{\partial A}{\partial P } \frac{\partial P}{\partial S }=2AW^\top P^{-2}WAS.
\end{align*}
Now we can calculate $\frac{\partial l}{\partial S}$:
\begin{align*}
\frac{\partial l}{\partial S} &=  -\frac{n\beta\beta^\top AW^\top P^{-2}WA S}{\tau^2+\beta^\top A\beta}\\
&+\frac{\beta\beta^\top AW^\top P^{-2}WA S}{2(\tau^2+\beta^\top A\beta)^2}\sum_{i=1}^n \left(r_i-\beta^\top AW^\top P^{-1}x_i\right)^2\\
    &+\frac{2}{(\tau^2+\beta^\top A\beta)} \sum_{i=1}^n \left(r_i-\beta^\top AW^\top P^{-1}x_i\right) \left(\beta^\top W^\top P^{-1}x_i  A W^\top P^{-2} W A S - \beta^\top AW^\top P^{-2} x_i S \right)\\
    &-nQ^{-1}S+\sum_{i=1}^nQ^{-1}x_ix_i^\top Q^{-1}S-mP^{-1}S+\sum_{j=1}^mP^{-1}y_jy_j^\top P^{-1}S.
\end{align*}
For $\frac{\partial l}{\partial W}$, we need to following building block:
\begin{align*}
\frac{\partial A}{\partial W } = -A(P^{-1}W+W^\top P^{-1})A
\end{align*}
Now we can calculate $\frac{\partial l}{\partial W}$:
\begin{align*}
\frac{\partial l}{\partial W} &=  \frac{n\beta\beta^\top A(P^{-1}W+W^\top P^{-1})A }{2(\tau^2+\beta^\top A\beta)}\\
&+\frac{\beta\beta^\top A(P^{-1}W+W^\top P^{-1})A}{2(\tau^2 +\beta^\top A\beta)^2}\sum_{i=1}^n \left(r_i-\beta^\top AW^\top P^{-1}x_i\right)^2\\
&-\frac{1}{(\tau^2+\beta^\top A\beta)} \sum_{i=1}^n \left(r_i-\beta^\top AW^\top P^{-1}x_i\right) \left(\beta^\top W^\top P^{-1}x_i A(P^{-1}W+W^\top P^{-1})A   - \beta^\top A P^{-1} x_i \right)\\
&-nQ^{-1}W+\sum_{i=1}^nQ^{-1}x_ix_i^\top Q^{-1}W.
\end{align*}

\begin{align*}
\frac{\partial l}{\partial\beta} &=   -\frac{nA\beta}{\tau^2+\beta^\top A\beta}+\frac{A\beta}{(\tau^2+\beta^\top A\beta)^2} \sum_{i=1}^n\left(r_i-\beta^\top AW^\top P^{-1}x_i\right)^2\\
&~~~~~+\frac{1}{2(\tau^2+\beta^\top A\beta)}\sum_{i=1}^n(r_i-\beta^\top AW^\top P^{-1}x_i)AW^\top P^{-1}x_i
\end{align*}

For $\frac{\partial l}{\partial \sigma^2}$, we need to following building block:
\begin{align*}
\frac{\partial P}{\partial \sigma^2 }=\frac{\partial Q}{\partial \sigma^2 }  = \Id,~\frac{\partial A}{\partial \sigma^2} = A^2W^\top P^{-2}W
\end{align*}

\begin{align*}
\frac{\partial l}{\partial\sigma^2} &=   -\frac{n\beta^\top A^2W^\top P^{-2}W\beta }{2(\tau^2+\beta^\top A\beta)}\\
&+\frac{\beta^\top A^2W^\top P^{-2}W\beta }{2(\tau^2 +\beta^\top A\beta)^2}\sum_{i=1}^n \left(r_i-\beta^\top AW^\top P^{-1}x_i\right)^2\\
&+\frac{1}{\tau^2 +\beta^\top A\beta}\sum_{i=1}^n \left(r_i-\beta^\top AW^\top P^{-1}x_i\right)\left(\beta^\top A^2W^\top P^{-2}W W^\top P^{-1}x_i-\beta^\top AW^\top P^{-2}x_i\right)\\
&-\frac{n}{2}\tr(Q^{-1}) +\frac{1}{2}\sum_{i=1}^n x_i^\top Q^{-2}x_i-\frac{m}{2} \tr(P^{-1})+\frac{1}{2}\sum_{j=1}^m y_j^\top P^{-2}y_j.
\end{align*}

Finally, we calculate $\frac{\partial l}{\partial \tau^2}$:
\begin{align*}
\frac{\partial l}{\partial \tau^2} &=   -\frac{n}{2(\tau^2+\beta^\top A\beta)}+\frac{1}{(\tau^2+\beta^\top A\beta)^2}\sum_{i=1}^n\left(r_i-\beta^\top AW^\top P^{-1}x_i\right)^2.
\end{align*}

\section{Proof of Theorem \ref{thm:predict}}
This is a direct consequence of Equation \eqref{eqn:r|x}.

\end{appendix}

\begin{acks}[Acknowledgments]
\end{acks}

\begin{funding}
Funding was provided in part by the Helmsley Trust grant AWD1006624, NIH NCI 5U2CCA233195, NSF CAREER AWD1005627, and CZI.  BEE is a CIFAR Fellow in the Multiscale Human Program. DL was supported by NIH grants R01 AG079291, R56 LM013784, R01 HL149683 and UM1 TR004406.
\end{funding}

\begin{supplement}
\begin{figure}[!ht]
    \centering
    \includegraphics[width=\textwidth]{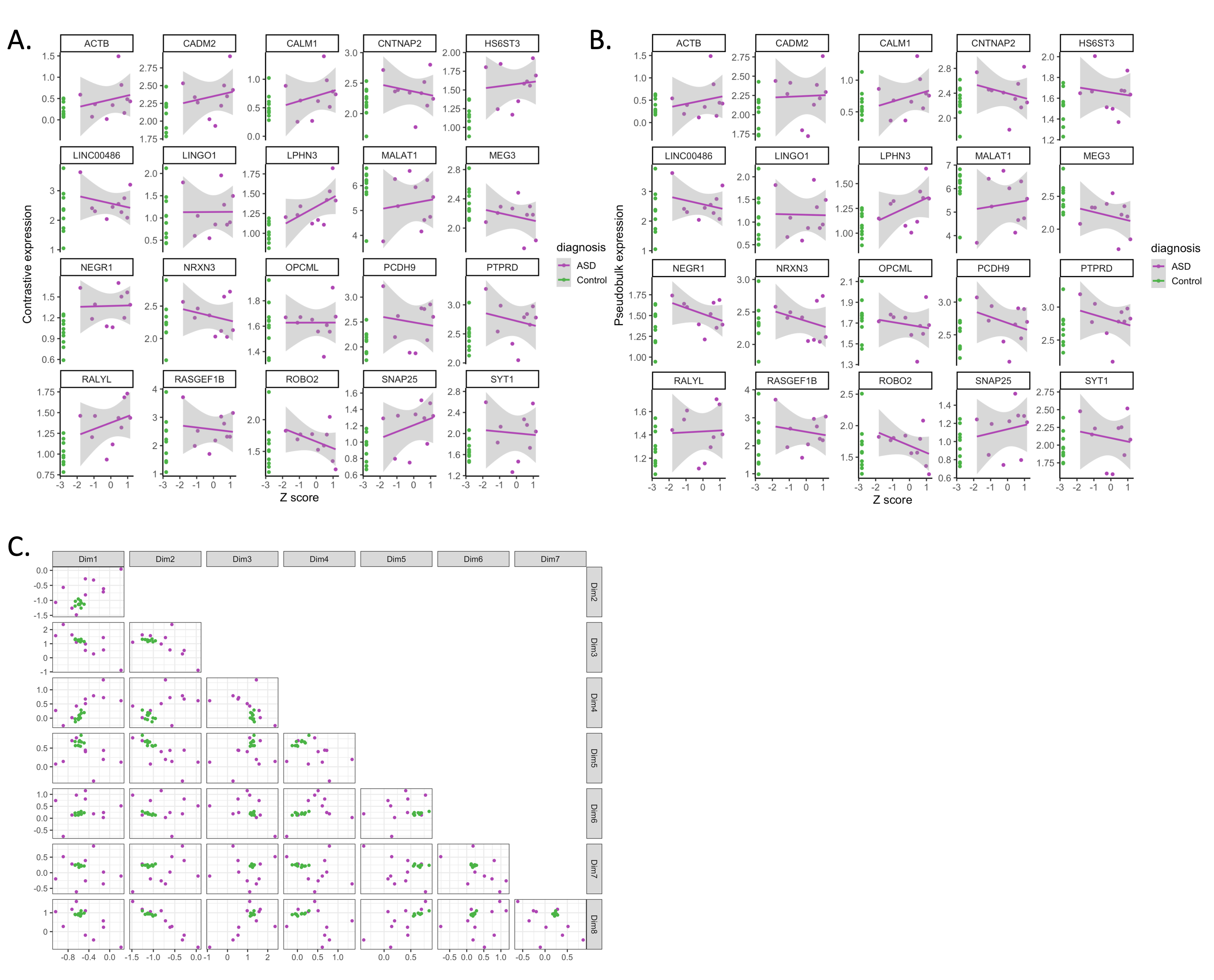}
    \label{fig:supp_asd_genes}
\end{figure}
\stitle{Supplementary Fig 1. Contrastive expression of top 20 genes and pairwise latent component in L2/3 excitatory neurons.}
\sdescription{Scatter plot of pseudobulk gene expression (A) and contrastive expression (B) against z-score for social behavior score for the top 20 genes identified by contrastive regression.  C. Pairwise scatter plot of latent dimensions in contrastive regression. Color indicates the diagnosis: green for controls and purple for autism.
}
\end{supplement}

\begin{supplement}
\begin{figure}[!ht]
    \centering
    \includegraphics[width=\textwidth]{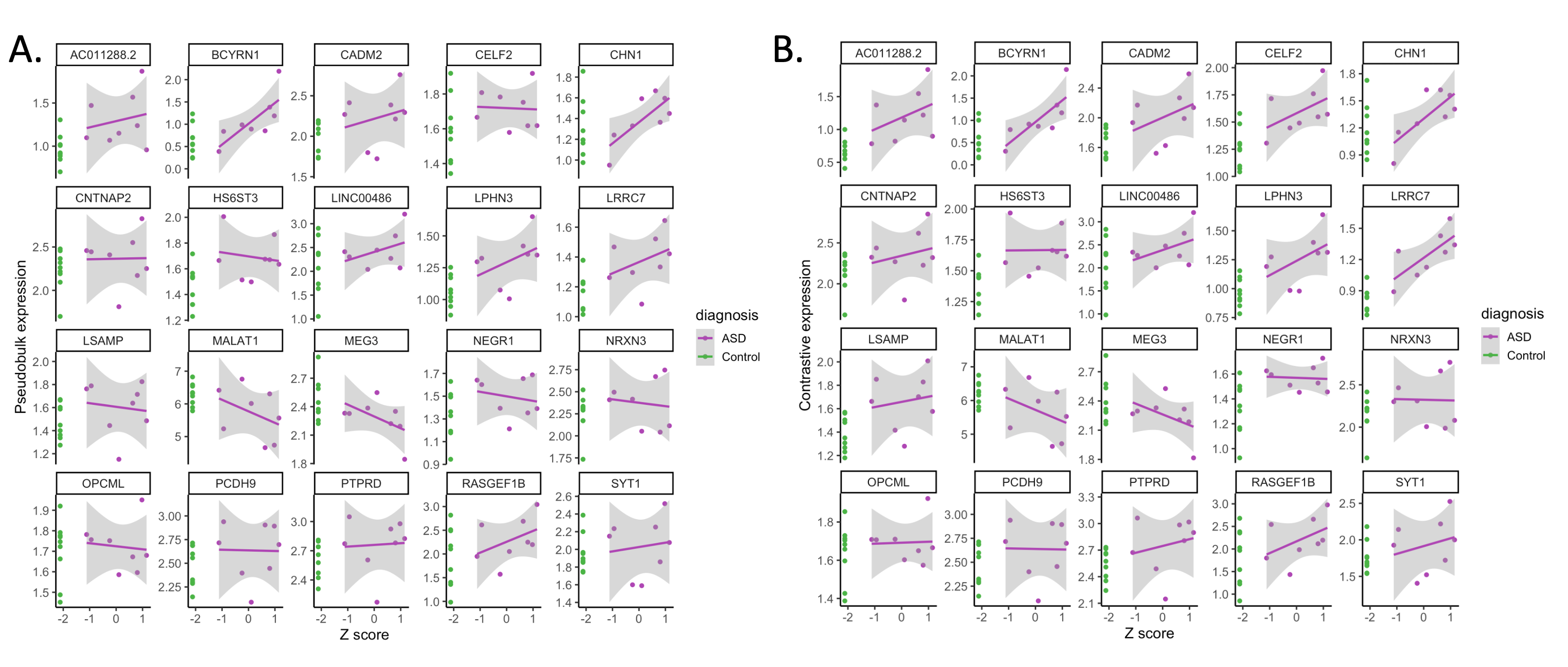}
    \label{fig:supp_sensitivity}
\end{figure}
\stitle{Supplementary Fig 2. Sensitivity analysis in L2/3 excitatory neurons.}
\sdescription{Scatter plot of pseudobulk gene expression (A) and contrastive expression (B) against z-score for social behavior score for the top 20 genes identified by contrastive regression, after excluding outliers.
}
\end{supplement}
\begin{supplement}
\begin{figure}[!ht]
    \centering
    \includegraphics[width=\textwidth]{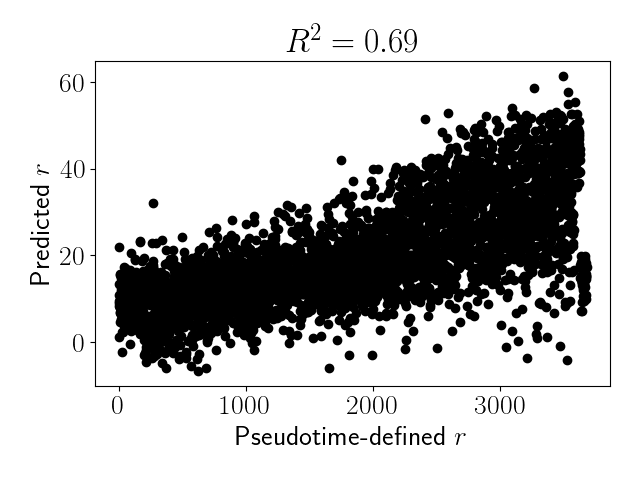}
    \label{fig:nopolyprsquared}
\end{figure}
\stitle{Supplementary Fig 3. Prediction of contrastive regression model on non-polyp data}
\sdescription{Scatter plot of predicted $r$ value compared to pseudotime ordering. Each point is a single cell from a non-polyp sample.
}
\end{supplement}

\begin{supplement}
\begin{figure}[ht]
    \centering
    \includegraphics[width=\textwidth]{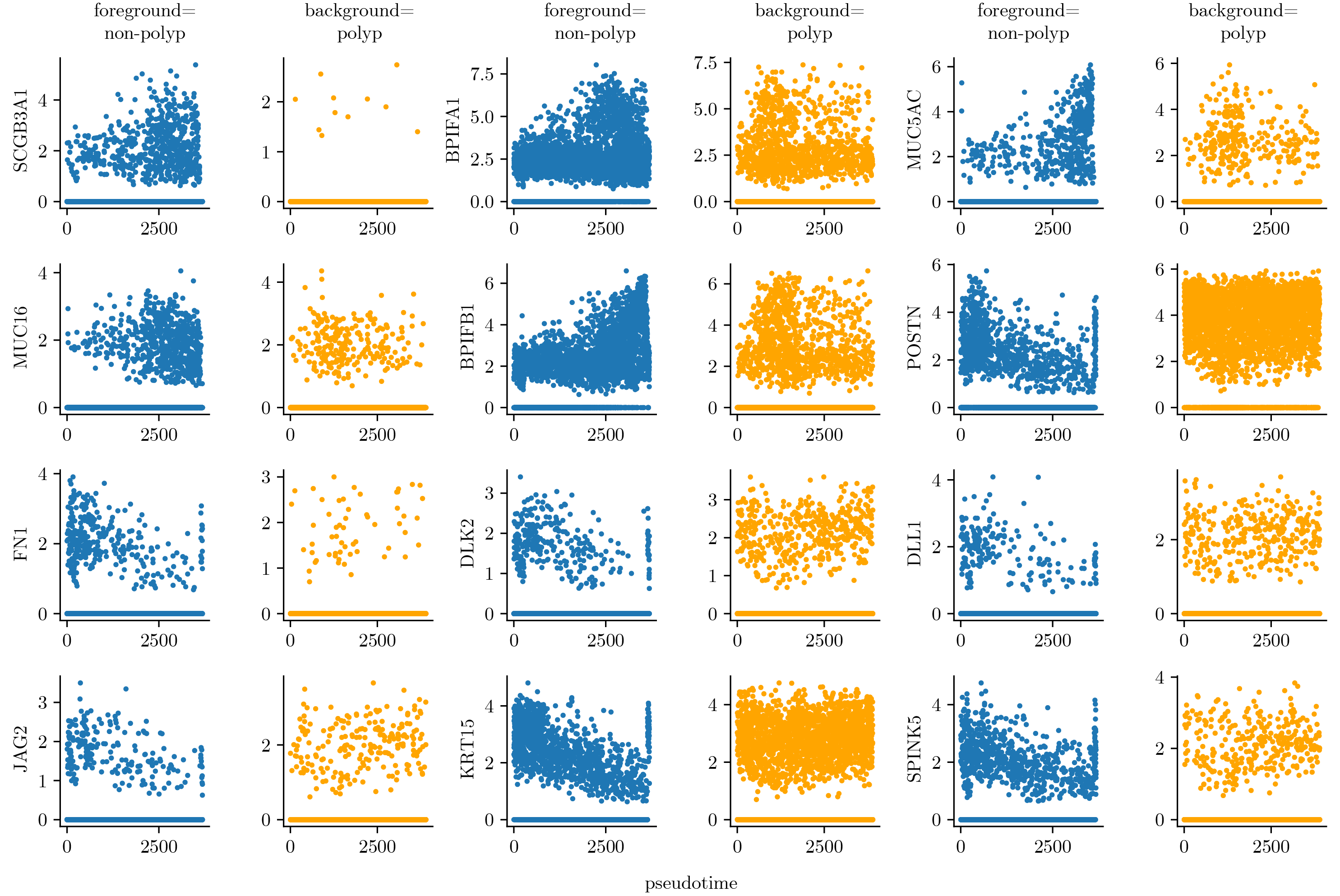}
    \label{fig:supp_polyp_genes}
\end{figure}
\stitle{Supplementary Fig 4. Comparison of gene expression in polyp and non-polyp cells}
\sdescription{Log10-normalized gene expression values of several genes identified via contrastive regression. Blue points show per-cell expression of each gene in cells from non-polyp (foreground) patients ordered by the pseudotime trajectory ordering (\textit{r}). Orange points show this gene expression in cells derived from donors with nasal polyps ordered along the trajectory and showing little correlation with trajectory location.
}
\end{supplement}
\newpage

\bibliographystyle{imsart-nameyear} 
\bibliography{ref}       
\end{document}